\documentclass[prd,aps,12pt,showpacs,nofootinbib,tightenlines]{revtex4}
\usepackage{amsmath}
\usepackage{amssymb}
\usepackage{epsfig}
\usepackage{graphicx}
\usepackage{bm}
\usepackage{times}
\usepackage{braket}
\usepackage{color}
\usepackage{slashed}
\usepackage{hyperref}

\usepackage{longtable}
\usepackage{comment}
\usepackage{subfigure}
\usepackage[]{units}
\usepackage{booktabs}
\usepackage{rotating}
\usepackage{dcolumn}
\definecolor{Red}{rgb}{1.,0.,0.}

\definecolor{Blue}{rgb}{0.,0.,1.}
\newcommand{\Blue}[1]{{\color{Blue}{#1}}}

\definecolor{nicered}{rgb}{0.7,0.1,0.1}
\definecolor{nicegreen}{rgb}{0.1,0.5,0.1}
\hypersetup{colorlinks,citecolor=nicegreen,linkcolor=nicered}

\newcommand{\beq}{\begin{eqnarray}}
\newcommand{\eeq}{\end{eqnarray}}

\newcommand{\calb}{ {\cal B} }

\def \epjc{ Eur. Phys. J. C }

\def \npb{  Nucl. Phys. B }
\def \plb{  Phys. Lett. B }

\def \prd{  Phys. Rev. D }
\def \prl{  Phys. Rev. Lett.  }

\def \jhep{ JHEP }

\begin{document}
\title{$S$, $P$ and $D$-wave resonance contributions to $B_{(s)} \to \eta_c(1S,2S) K\pi$ decays in the perturbative QCD approach}
\author{Ya Li$^1$}                \email[]{liyakelly@163.com}
\author{Da-Cheng Yan$^2$}         \email[]{yandac@126.com}
\author{Zhou Rui$^3$}              \email[]{jindui1127@126.com}
\author{Zhen-Jun Xiao$^{4}$}    \email[]{xiaozhenjun@njnu.edu.cn}

\affiliation{$^1$ Department of Physics, College of Sciences, Nanjing Agricultural University, Nanjing, Jiangsu 210095, P.R. China}
\affiliation{$^2$ School of Mathematics and Physics, Changzhou University, Changzhou, Jiangsu 213164, P.R. China}
\affiliation{$^3$ College of Sciences, North China University of Science and Technology,
                          Tangshan 063009,  P.R. China}
\affiliation{$^4$ Department of Physics and Institute of Theoretical Physics,
                          Nanjing Normal University, Nanjing, Jiangsu 210023, P.R. China}

\date{\today}

\begin{abstract}
In this work, we analyze the three-body $B_{(s)} \to \eta_c(1S,2S) K \pi$ decays within the framework of the perturbative QCD approach (PQCD) under the quasi-two-body approximation, where the kaon-pion invariant mass spectra are dominated by the $K_0^*(1430)^0,K_0^*(1950)^0,K^*(892)^0,K^*(1410)^0,K^*(1680)^0$ and $K_2^*(1430)^0$ resonances.
The time-like form factors are adopted to parametrize the corresponding $S$, $P$, $D$-wave kaon-pion distribution amplitudes for the concerned decay modes, which describe the final-state interactions between the kaon and pion in the resonant region.
The $K\pi$ $S$-wave component at low $K\pi$ mass is described by the LASS line shape, while the time-like form factors of other resonances are modeled by the relativistic Breit-Wigner function.
We find the following main points:
(a) the PQCD predictions of the branching ratios for most considered $B \to \eta_c(1S)(K^{*0}\to )K^+\pi^-$ decays agree well with the currently available data within errors;
(b) for ${\cal B}(B^0 \to \eta_c (K_0^*(1430)\to )K^+\pi^-)$
and ${\cal B}(B^0 \to \eta_c K^+\pi^-({\rm NR}))$ (here NR means nonresonant), our predictions of the branching ratios are a bit smaller than the measured ones;
and (c) the PQCD results for the $D$-wave contributions considered in this work can be tested once the precise data from the future
 LHCb and Belle-II experiments are available.

\end{abstract}


\maketitle


\section{Introduction}
The $B_{(s)}$ meson decays into charmonia and a kaon-pion pair are of great interest since only a few color-suppressed modes in hadronic $B$ decays
have been measured so far.  Some standard model (SM) parameters can be extracted from the $b\to c\bar{c} s$  transitions, while the studies of these decay channels can also provide
an ideal place to find a signal  for the physics beyond the SM.
The meson $\eta_c$ and $J/\psi$ have the same quark content but with different spin angular momentum.
As expected, the $B$ meson decays involving the $\eta_c$ will garner considerable experimental attention with the development of experiments.
In recent years, significant improvements in understanding the heavy quarkonium production mechanism have been achieved~\cite{epjc75-311}.
The $B^0 \to \eta_c (K^*(892)^0\to) K\pi$ decay has been observed by Belle~\cite{Prl90-071801} and {\it BABAR}~\cite{PRD76-092004,PRD78-012006} collaborations.
Very recently, the $B^0 \to \eta_c K^+\pi^-$ decay  was measured for the first time by the LHCb Collaboration~\cite{epjc78-1019},
with the $\eta_c$ meson reconstructed using the $p\bar{p}$ decay mode.
This  decay is expected to proceed through $K^{*0} \to K\pi$ intermediate states as well as the nonresonant (NR) S-wave component,
where the $K^{*0}$ refers to various partial wave resonances, such as $K_0^*(1430)^0,K_0^*(1950)^0,K^*(892)^0,K^*(1410)^0,K^*(1680)^0$, and $K_2^*(1430)^0$.
The $P$-wave $K^*(892)^0$ is the largest component, $\sim 50\%$, while the $K^*(1410)^0,K^*(1680)^0$, and $D$-wave $K_2^*(1430)^0$ states amount
to only a few percent.

The theoretical study and experimental measurement of the three-body $B$ meson decays is still in an early stage.
On the theoretical side, compared with those two-body decay modes,  these three-body $B$ decays are less tractable due to the entangled resonant
and nonresonant contributions, as well as the possible final-state interactions~\cite{prd89-094013,1512-09284,prd89-053015}.
An important breakthrough in the theory of three-body $B$ meson decays was the confirmation of the validity of factorization.
We restrict ourselves to the specific kinematical configurations in which the three mesons are quasialigned in the rest frame of the $B$ meson.
This condition is particularly natural in the low-effective-$K\pi$-mass region of the Dalitz plot, where most of the $K\pi$ resonant structures are seen.
When the two particles among the three final-state mesons move collinearly and generate a small invariant mass recoiling against the third one,
the three-body interactions are expected to be suppressed.
Then, it seems reasonable to assume the validity of the factorization for these quasi-two-body $B$ decays.
In the quasi-two-body mechanism, the two-body scattering and all possible interactions between the two involved particles are included, but the interactions
between the third particle and the pair of mesons are ignored.
In recent years, several different theoretical frameworks based on the factorization theorems and symmetry principles have been proposed to deal with the three-body $B$ meson decays.
The QCD-improved factorization approach~\cite{prl83-1914,npb591-313,npb606-245,npb675-333} has been widely used
in the study of the three-body charmless hadronic $B$ meson
decays~\cite{npb899-247,plb622-207,prd74-114009,prd79-094005,APPB42-2013,prd76-094006,prd88-114014,prd94-094015,
prd89-094007,prd87-076007,jhep10-117}.
The $U$-spin and the flavor $SU$(3) symmetries were also adopted in
Refs.~\cite{prd72-094031,plb727-136,prd72-075013,prd84-056002,plb728-579,prd91-014029}.

It has been known that the collinear factorization of the charmed and charmless two-body $B$ meson decays suffer from end-point singularities.
The perturbative QCD(PQCD) factorization approach relying on the $k_T$ factorization theorem \cite{plb561-258,prd70-054006} was proposed in Refs.~\cite{prd55-5577,plb504-6,prd63-074009},
which has been shown to be infrared finite, gauge invariant, and consistent with the factorization assumption in the heavy-quark
limit~\cite{prl74-4388,prd53-2480,prd65-014007,plb555-197}.
The operator-level definition of the transverse-momentum-dependent hadronic wave functions is highly nontrivial in order to avoid
the potential light-cone divergence and the rapidity singularity~\cite{JHEP06-013,JHEP02-008}.
The Sudakov factors from the $k_T$ resummation have been included to suppress the long-distance contributions from the large-$b$
region with $b$ being a variable conjugate to $k_T$.
Therefore, the PQCD approach is a self-consistent framework and has good predictive power.
Based on the PQCD approach, the quasi-two-body $B$ meson decays have been studied in Refs.~\cite{plb763-29,epjc76-675,prd95-056008,npb924-745,prd98-056019,prd98-113003,prd97-034033,epjc79-37,plb791-342,epjc79-539,epjc79-792,1907-10422}
by introducing the two-meson distribution amplitudes (DAs)~\cite{fp42-101,prl81-1782,prd62-073014,
epjc26-261,npb555-231,sjnp38-289,tmp69-1109}, which catch the dynamics associated with the pair of mesons.

\begin{figure}[tbp]
\centerline{\epsfxsize=16cm \epsffile{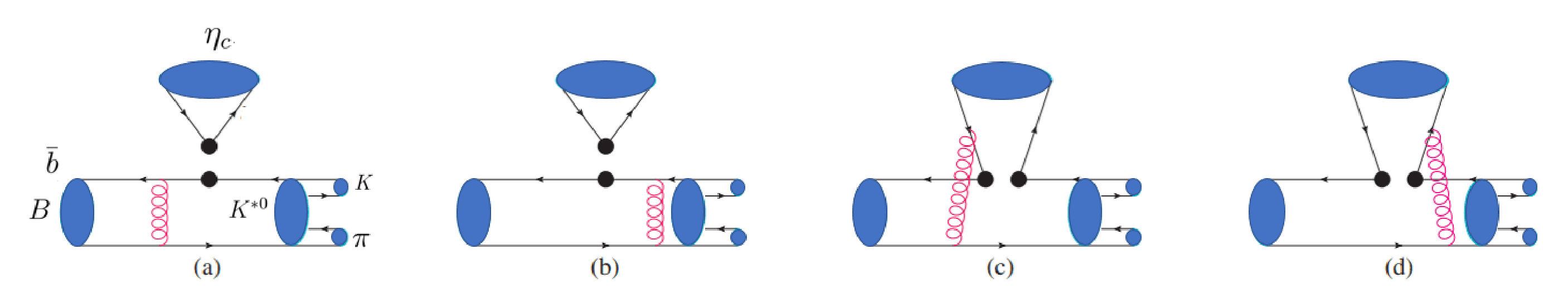}}
\caption{Typical diagrams for the quasi-two-body decays $B \to \eta_c{(1S,2S)}(K^{*0} \to) K \pi$, where the symbol (Bullet) denotes the weak vertex.
$K^{*0}$ represents various partial-wave intermediate states.}
\label{fig:fig1}
\end{figure}

In this paper, we continue to study the quasi-two-body decays $B\to \eta_c K^{*0} \to \eta_c K\pi$  involving the $S$, $P$, and $D$-wave kaon-pion pairs as shown in Fig.~\ref{fig:fig1}, within the framework of the PQCD factorization approach.
Some studies of $B \to \eta_c K^*$ decays have used the two-body framework~\cite{prd71-114008,1909-10907,epjc58-245}.
From Refs.~\cite{plb763-29,prd95-056008,prd98-056019}, we know that the width of the resonant state and the interactions between the final-state meson pair will
show their effects on the branching ratios, especially on the direct $CP$ violations of the quasi-two-body decays.
Thus, it seems more appropriate to treat the $K^{*0}$ as an intermediate resonance.
As addressed before, this process is dominated by a series of resonances in $S$, $P$, and $D$ waves.
The $S$-wave kaon-pion DAs for the resonance $K_0^*(1430)^0$ have been studied in Ref.~\cite{prd97-033006}, and we will further investigate the dependence of the branching ratios in different scalar scenarios as proposed in Refs.~\cite{Gorishnii:1983zi,Kataev:2004ve,prd73-014017,prd77-014034}.
Besides, we have roughly determined the possible range of the first odd Gegenbauer moment $B_1$ for the $K_0^*(1950)^0$ resonance by fitting to the
existing data, which must be tested in the future.
We intend to adopt the same fitted parameters as those of the longitudinal kaon-pion DAs in Ref.~\cite{1907-10422}, where the $SU$(3)
flavor-symmetry-breaking effect has been considered and plays an important role in the longitudinal polarization fractions.
The $D$-wave resonance $K_2^*(1430)^0$ is investigated for the first time in our work.
Due to the limited studies on the tensor resonant states, we treat the $D$-wave DAs of the $K_2^*(1430)^0$ in the same way as those of $f_2(1270)$~\cite{prd98-056019}.

This paper is organized as follows. In Sec. II we give a brief introduction of the theoretical framework.
The numerical values, some discussions, and our conclusions will be given in last two sections.
The explicit PQCD factorization formulas for all of the decay amplitudes are collected in the Appendix.

\section{Framework}\label{sec:2}  

In the framework of the PQCD factorization approach, the nonperturbative dynamics associated with the pair
of mesons can be absorbed into two-meson DAs; then, the relevant decay amplitude $\mathcal{A}$ for the quasi-two-body
decays $B\to \eta_c K^{*0} \to \eta_c K\pi$ can be written in the following form:
\begin{eqnarray}
\mathcal{A}=\Phi_B\otimes H\otimes \Phi_{K\pi}\otimes\Phi_{\eta_c},
\end{eqnarray}
where $\Phi_B$ and $\Phi_{\eta_c}$ are the $B$ meson and charmonium DAs, respectively.
The kaon-pion DA $\Phi_{K\pi}$ absorbs the nonperturbative dynamics in the $K\pi$ hadronization process.
The hard kernel $H$ contains only one hard gluon and describes the dynamics of the strong and electroweak interactions in the three-body hadronic decays as in the formalism for the corresponding two-body decays.

In the light-cone coordinates, we make the kaon-pion pair and the final-state $\eta_c$ move along the directions $n=(1,0,0_{\rm T})$ and $v=(0,1,0_{\rm T})$, respectively.
The $B$ meson momentum $p_{B}$, the total momentum of the kaon-pion pair, $p=p_1+p_2$, the final-state $\eta_c$ momentum $p_3$,  and the quark momentum $k_i$ in each meson are
\begin{eqnarray}
p_{B}&=&\frac{m_{B}}{\sqrt 2}(1,1,0_{\rm T}), ~\quad k_{B}=\left(0,x_B \frac{m_{B}}{\sqrt2} ,\textbf{k}_{B \rm T}\right),\nonumber\\
p&=&\frac{m_{B}}{\sqrt2}(1-r^2,\eta,0_{\rm T}), ~\quad k= \left( z (1-r^2)\frac{m_{B}}{\sqrt2},0,\textbf{k}_{\rm T}\right),\nonumber\\
p_3&=&\frac{m_{B}}{\sqrt 2}(r^2,1-\eta,0_{\rm T}), ~\quad k_3=\left(r^2x_3\frac{m_B}{\sqrt{2}},(1-\eta)x_3 \frac{m_B}{\sqrt{2}},\textbf{k}_{3{\rm T}}\right),\label{mom-B-k}
\end{eqnarray}
where $m_{B}$ is the mass of the $B$ meson, $\eta=\frac{\omega^2}{m^2_{B}(1-r^2)}$ with $r=m_{\eta_c}/m_{B}$, $m_{\eta_c}$ is the mass of charmonia, and the invariant mass squared $\omega^2=p^2$.
The momentum fractions $x_{B}$, $z$, and $x_3$ run from zero to unity, respectively.

We also define the momenta $p_1$ and $p_2$ in the kaon-pion pair as
\begin{eqnarray}\label{eq:p1p2}
 p_1&=&(\zeta p^+, (1-\zeta)\eta p^+, \textbf{p}_{1\rm T}), ~\quad p_2=((1-\zeta) p^+, \zeta\eta p^+, \textbf{p}_{2\rm T}),
\end{eqnarray}
with $\zeta=p_1^+/P^+$ characterizing the distribution of the longitudinal momentum of the kaon and $p_{1\rm T}^2=p_{2\rm T}^2=\zeta(1-\zeta)\omega^2$.

The $B_{(s)}$ meson wave function and the charmonium DAs are the same as widely adopted in the PQCD approach~\cite{epjc76-675,npb924-745,epjc79-792,epjc75-293}.
Very recently, a new method was proposed to calculate the $B$ meson light-cone DA from lattice QCD, which can be used as an updated input for the $B$ meson DA in the future~\cite{1908-09933}.
Below, we briefly describe the $S$, $P$, and $D$-wave kaon-pion DAs and the corresponding time-like form factors.
The $S$-wave kaon-pion DAs are introduced  analogously  to the case of two-pion ones~\cite{prd91-094024},
\begin{eqnarray}\label{eq:fuliye2}
\Phi_{S}=\frac{1}{\sqrt{2N_c}}[{p\hspace{-2.0truemm}/}\phi^0_S(z,\zeta,\omega^2)+
\omega\phi^s_S(z,\zeta,\omega^2)+\omega({n\hspace{-2.0truemm}/}{v\hspace{-2.0truemm}/}-1)\phi^t_S(z,\zeta,\omega^2)].
\end{eqnarray}
In what follows the subscripts $S$, $P$, and $D$ are always associated with the corresponding partial waves.

We will use the asymptotic forms for the twist-3 DAs, but no knowledge on the twist-2 DAs is available at present.
We shall adopt similar formulas as those for a scalar meson~\cite{plb730-336,prd73-014017}, bearing in mind large uncertainties that may be introduced by this approximation.
The detailed expressions of DAs of various twists are as follows:
\begin{eqnarray}\label{eq:phi0st}
\phi^0_S(z,\zeta,\omega^2)&=&\frac{6}{2\sqrt{2N_c}}F_S(\omega^2)z(1-z)
\left [ \frac{1}{\mu_S}+B_1C_1^{3/2}(t)+B_3C_3^{3/2}(t) \right ], \label{eq:phi1s}\\
\phi^s_S(z,\zeta,\omega^2)&=&\frac{1}{2\sqrt{2N_c}}F_S(\omega^2),\label{eq:phi2s}\\
\phi^t_S(z,\zeta,\omega^2)&=&\frac{1}{2\sqrt{2N_c}}F_S(\omega^2)(1-2z)\label{eq:phi3s},
\end{eqnarray}
where the Gegenbauer polynomials $C_{1}^{3/2}(t)=3t$, $C_{3}^{3/2}t=\frac{5}{2}t(7t^2-3)$ with $t=1-2z$ and $\mu_S=\omega/(m_2-m_1)$.
The Gegenbauer moments $B_{1,3}$ and the related running current quark masses can be found in Refs.~\cite{prd73-014017,prd77-014034,prd78-014006}.
It should be stressed that there is less information about the scalar resonance $K_0^*(1950)^0$, and we only test the sensitivity of the branching ratios on the first odd Gegenbauer moment $B_1$ in our work.

If there are overlapping resonances or there is significant interference with a nonresonant component both in the same partial wave, the relativistic Breit-Wigner (RBW) function leads to unitarity violation within the isobar model~\cite{0712-1605}.
This is the case for the $K\pi$ $S$-wave  at low $K\pi$ mass, where the $K^*_0(1430)^0$ resonance interferes strongly with a slowly varying NR $S$-wave component.
In this work, the time-like scalar form factor $F_S(\omega^2)$ for the $S$-wave $K\pi$ system is parametrized by using a modified LASS line shape~\cite{npb296-493} for the $S$-wave resonance $K^*_0(1430)^0$, which has been widely used in experimental analyses~\cite{epjc78-1019},
\begin{eqnarray}\label{eq:lass}
F_S(\omega^2)=\frac{\omega}{|\vec{p}_1|[\cot(\delta_B)-i]}+e^{2i\delta_B}\frac{m_0^2
\Gamma_0/|\vec{p}_{0}|}{m_0^2-\omega^2-im_0^2 \frac{\Gamma_0}{\omega }\frac{|\vec{p}_1|}{|\vec{p}_{0}|}}\;,
\end{eqnarray}
with
\begin{eqnarray}\label{eq:ar}
\cot(\delta_B)=\frac{1}{a|\vec{p}_1|}+\frac{r|\vec{p}_1|}{2},
\end{eqnarray}
where the first term in Eq.~(\ref{eq:lass}) is an empirical term from the elastic kaon-pion scattering and the second term is the resonant contribution with a phase factor to retain unitarity.
Here $m_0$ and $\Gamma_0$ are the pole mass and width of the $K^*_0(1430)$ state.
$|\overrightarrow{p_1}|$ is the momentum vector of the resonance decay product measured in the resonance rest frame, and $|\overrightarrow{p_0}|$
is the value of $|\overrightarrow{p_1}|$ when $\omega=m_{K^*}$.
The parameters $a=(3.1\pm 1.0)~{\rm GeV^{-1}}$ and  $r=(7.0\pm 2.4)~{\rm GeV^{-1}}$ are the scattering length and effective range~\cite{epjc78-1019}, respectively,
which are universal in application to the description of different processes involving the kaon-pion pair.
The slowly varying part [the first term in the Eq.~(\ref{eq:lass})] is not well modeled at high masses and it is set to  zero for $m(K\pi)$ values above 1.7 GeV~\cite{epjc78-1019}.
For the $K_0^*(1950)^0$, we use the relativistic BW line shape to  parametrize the time-like
form factors $F_S(\omega^2)$, which is adopted in the experimental data analyses~\cite{epjc78-1019}.

The $P$-wave kaon-pion DAs related to both longitudinal and transverse polarizations have been studied in Ref.~\cite{1907-10422}.
In quasi-two-body $B \to \eta_c K^*(892)^0\to \eta_c K\pi$ decay, the vector meson $K^*(892)$ should be completely polarized in the longitudinal direction due to the angular momentum conservation requirement.
The explicit expressions of the $P$-wave kaon-pion DAs associated with longitudinal polarization are listed as follows,
\begin{eqnarray}\label{eq:pwavekpi}
\Phi_{P}&=&\frac{1}{\sqrt{2N_c}}
\left [{p\hspace{-2.0truemm}/}\phi^0_P(z,\zeta,\omega^2)+\omega \phi^s_P(z,\zeta,\omega^2)+
\frac{{p\hspace{-1.5truemm}/}_1{p\hspace{-1.5truemm}/}_2-{p\hspace{-1.5truemm}/}_2{p\hspace{-1.5truemm}/}_1}{\omega(2\zeta-1)}
\phi^t_P(z,\zeta,\omega^2) \right ]\Blue{.}
\end{eqnarray}
The DAs of various twists in Eq.~(\ref{eq:pwavekpi}) can be expanded in terms of the Gegenbauer polynomials:
\begin{eqnarray}
\label{ptst}
\phi_P^0(z,\zeta,\omega^2)&=&\frac{3F_{P}^{\parallel}(\omega^2)}{\sqrt{2N_c}} z(1-z)\left[1+a_{1K^*}^{||}3t+a_{2K^*}^{||}\frac{3}{2}(5t^2-1)\right](2\zeta-1-\alpha)\;,\label{eq:phi1p}\\
\phi_P^s(z,\zeta,\omega^2)&=&\frac{3F_{P}^{\perp}(\omega^2)}{2\sqrt{2N_c}}\bigg\{t\left[1+a_{1s}^{\perp}t\right]-a_{1s}^{\perp}2z(1-z)\bigg\}P_1(2\zeta-1) \;,\label{eq:phi2p}\\
\phi_P^t(z,\zeta,\omega^2)&=&\frac{3F_{P}^{\perp}(\omega^2)}{2\sqrt{2N_c}}t\left[t+a_{1t}^{\perp}(3t^2-1)\right]P_1(2\zeta-1) \;,\label{eq:phi3p}
\end{eqnarray}
where the Legendre polynomial $P_1(2\zeta-1)=2\zeta-1$ and the factor $\alpha=(m^2_{K}-m^2_{\pi})/\omega^2$ is treated as the $SU$(3) asymmetry factor.

The Gegenbauer moments $a_{1K^*}^{||},a_{2K^*}^{||},a_{1s}^{\perp},a_{1t}^{\perp}$ are adopted the same
values as those determined in Ref.~\cite{1907-10422}:
\begin{eqnarray}
a_{1K^*}^{\parallel}&=&0.2, ~\quad a_{2K^*}^{\parallel}=0.5,~\quad a_{1s}^{\perp}=-0.2,~\quad a_{1t}^{\perp}=0.2.
\end{eqnarray}

The relativistic BW line shape is adopted for the $P$-wave resonances $K^*(892)$, $K^*(1410)$, and $K^*(1680)$ to  parametrize the time-like form factors $F_{P}^{\parallel}(\omega^2)$.
The explicit expression is~\cite{epjc78-1019},
\begin{eqnarray}
\label{BRW}
F_{P}^{\parallel}(\omega^2)&=&\frac{c_1 m_{K^*(892)}^2}{m^2_{K^*(892)} -\omega^2-im_{K^*(892)}\Gamma_1(\omega^2)}+\frac{c_2 m_{K^*(1410)}^2}{m^2_{K^*(1410)} -\omega^2-im_{K^*(1410)}\Gamma_2(\omega^2)}\nonumber\\
&+&\frac{c_3 m_{K^*(1680)}^2}{m^2_{K^*(1680)} -\omega^2-im_{K^*(1680)}\Gamma_3(\omega^2)}\;,
\end{eqnarray}
where the three terms describe the contributions from $K^*(892)$, $K^*(1410)$, and $K^*(1680)$, respectively.
The weight coefficients $c_1=0.72, c_2=0.135, c_3=0.145$ are the same as those being determined previously~\cite{1907-10422}.

The mass-dependent width $\Gamma_i(\omega)$ is defined as
\begin{eqnarray}
\label{BRWl}
\Gamma_i(\omega^2)&=&\Gamma_i\left(\frac{m_i}{\omega}\right)\left(\frac{|\overrightarrow{p_1}|}{|\overrightarrow{p_0}|}\right)^{(2L_R+1)}\Blue{.}
\end{eqnarray}
The $m_i$ and $\Gamma_i$ are the pole mass and width of the corresponding resonances, where $i=1,2,3$ represents the resonances $K^*(892)$, $K^*(1410)$, and $K^*(1680)$, respectively.
$L_R$ is the orbital angular momentum in the $K^+\pi^-$ system and $L_R=0,1,2,...$ corresponds to the $S,P,D,...$ partial-wave resonances.
Following Ref.~\cite{plb763-29}, we also assume that
\begin{eqnarray}
F_{P}^{\perp}(\omega^2)/F_{P}^{\parallel}(\omega^2)\approx (f_{K^*}^T/f_{K^*}),
\end{eqnarray}
with $f_{K^*}=0.217 \pm 0.005$ {\rm GeV} and $f^T_{K^*}=0.185 \pm 0.010$ {\rm GeV}~\cite{prd76-074018}.
Due to the limited studies on the decay constants of $K^*(1410)$ and $K^*(1680)$, we use the two decay
constants of $K^*(892)$ to determine the ratio $F_{P}^{\perp}(\omega^2)/F_{P}^{\parallel}(\omega^2)$.

The $D$-wave kaon-pion DAs are introduced analogously to the two-pion ones~\cite{prd98-056019},
\begin{eqnarray}\label{eq:dwavekpi}
\Phi_{D}&=&\frac{1}{\sqrt{2N_c}}
\left [ {p\hspace{-2.0truemm}/}\phi^0_D(z,\zeta,\omega^2)+\omega \phi^s_D(z,\zeta,\omega^2)+
\frac{{p\hspace{-1.5truemm}/}_1{p\hspace{-1.5truemm}/}_2-{p\hspace{-1.5truemm}/}_2{p\hspace{-1.5truemm}/}_1}{\omega(2\zeta-1)}
\phi^t_D(z,\zeta,\omega^2) \right ].
\end{eqnarray}
The $D$-wave $K\pi$ system has similar asymptotic DAs as those for a tensor
meson~\cite{prd83-034001,prd83-014008,prd86-094015}, but with the time-like form factor replacing the tensor decay constants:
\begin{eqnarray}\label{eq:dwavetwist3}
\phi^0_D(z,\zeta,\omega^2)&=&\frac{6F^{\parallel}_D(\omega^2)}{2\sqrt{2N_c}}z(1-z) \left [3a_1^0(2z-1) \right ] P_2(2\zeta-1),\label{eq:phi1d}\\
\phi^s_D(z,\zeta,\omega^2)&=&-\frac{9F^{\perp}_D(\omega^2)}{4\sqrt{2N_c}}\left [ a_1^0(1-6z+6z^2) \right ] P_2(2\zeta-1),\label{eq:phi2d}\\
\phi^t_D(z,\zeta,\omega^2)&=&\frac{9F^{\perp}_D(\omega^2)}{4\sqrt{2N_c}}\left [ a_1^0(1-6z+6z^2)(2z-1) \right ] P_2(2\zeta-1).\label{eq:phi3d}
\end{eqnarray}
where the Legendre polynomial $P_2(2\zeta-1)=1-6\zeta(1-\zeta)$ and the Gegenbauer moment $a_1^0=0.4\pm0.1$~\cite{prd98-056019}.
The time-like form factor $F^{\parallel}_D(\omega^2)$ for the $D$-wave $K\pi$ resonance is also described by the relativistic BW function as given in Eq.~(\ref{BRW}).
Besides, the approximate relation $F_{D}^{\perp}(\omega^2)/ F_{D}^{\parallel}(\omega^2)\approx f_{K_2^*(1430)}^T/f_{K_2^*(1430)}$
is used in the following calculation with $f_{K_2^*(1430)}=0.118 \pm 0.005$ {\rm GeV} and $f^T_{K_2^*(1430)}=0.077 \pm 0.014$ {\rm GeV}~\cite{prd83-034001}.

\section{Numerical results and discussions}\label{sec:3}  

\begin{table}  
\caption{ The pole masses and widths of the various partial-wave resonances~\cite{epjc78-1019}.}
\label{Tab:pa}
\begin{tabular*}{10cm}{@{\extracolsep{\fill}}lllll} \hline\hline
{\rm Resonance} &Mass~[MeV] &Width~[MeV] &$J^{P}$&Model\\ \hline
$K^*(892)^0$      &$895.55\pm0.20$&$47.3\pm0.5$&$1^{-}$&\rm RBW\\
$K^*(1410)^0$      &$1414\pm 15$&$232\pm21$&$1^{-}$&\rm RBW\\
$K_0^*(1430)^0$    &$1425\pm 50$&$270\pm80$&$0^{+}$&\rm LASS\\
$K_2^*(1430)^0$    &$1432.4\pm 1.3$&$109\pm5$&$2^{+}$&\rm RBW\\
$K^*(1680)^0$      &$1717\pm27$&$322\pm110$ &$1^{-}$&\rm RBW\\
$K_0^*(1950)^0$    &$1945\pm 22$&$201\pm90$&$0^{+}$&\rm RBW\\
\hline\hline
\end{tabular*}
\end{table}

In our numerical calculations, besides the quantities specified before, the following input parameters (where the masses and decay constants are in units of {\rm GeV} ) are used~\cite{pdg2018}:
\begin{eqnarray}
m_{B^{0}}&=&5.28, \quad m_{B_s^0}=5.367, \quad m_{b}=4.8, \quad m_c=1.275,\nonumber\\
m_{\pi^\pm}&=&0.140, \quad m_{K^{\pm}}=0.494, \quad m_{\eta_c(1S)}=2.9834, \quad m_{\eta_c(2S)}=3.6392, \nonumber\\
f_{B^0}&=&0.19, \quad f_{B_s}=0.23,\quad f_{\eta_c}= 0.42,\quad f_{\eta_c(2S)}= 0.243,\nonumber\\
\tau_{B^0}&=&1.519\; {\rm ps}, \quad \tau_{B^0_s}=1.512\; {\rm ps}.\label{eq:inputs}
\end{eqnarray}
The pole masses and widths of various partial-wave resonances are summarized in Table~\ref{Tab:pa},
while we adopt the values of the Wolfenstein parameters  given in
Ref.~\cite{pdg2018}: $A=0.836\pm0.015, \lambda=0.22453\pm 0.00044$, $\bar{\rho} = 0.122^{+0.018}_{-0.017}$, $\bar{\eta}= 0.355^{+0.012}_{-0.011}$.

For the decays $B \to \eta_c (K^{*0} \to) K \pi$, the differential decay rate is expressed as
\begin{eqnarray}
\frac{d{\cal B}}{d\omega}=\frac{\tau_{B}\omega|\overrightarrow{p_1}||\overrightarrow{p_3}|}{32\pi^3m^3_{B}}|{\cal A}|^2, \label{expr-br}
\end{eqnarray}
where the kaon and charmonium three-momenta in the $K\pi$ center-of-mass frame are given by
\begin{eqnarray}
|\overrightarrow{p_1}|=\frac{1}{  2\omega}      \sqrt{\lambda(\omega^2,m_K^2,m^2_{\pi})} , \quad~~
|\overrightarrow{p_3}|=\frac{1}{2\omega} \sqrt{\lambda(m_B^2,m^2_{\eta_c},\omega^2) } ,
 \label{br-momentum}
\end{eqnarray}
with the kaon (pion) mass $m_K$ ($m_{\pi})$ and the K$\ddot{a}$ll$\acute{e}$n function $\lambda (a,b,c)= a^2+b^2+c^2-2(ab+ac+bc)$.

\begin{table}  
\caption{ PQCD results for the branching ratios  of the $S$-wave resonances in the
$B^0_{(s)} \to \eta_c(1S,2S) K^{\pm} \pi^{\mp}$ decays in scenario I and scenario II together with experimental data~\cite{epjc78-1019}.
The theoretical errors are attributed to the variation of the Gegenbauer moments $B_1$ and $B_3$, the shape parameters $\omega_{B_{(s)}}$
in the wave function of the $B_{(s)}$
meson, and the hard scale $t$, respectively. }
\label{Tab:swave}
\begin{center}
\begin{tabular*}{16cm}{@{\extracolsep{\fill}}l l l l}
\hline\hline
\multicolumn{1}{c}{}&\multicolumn{2}{c}{ Quasi-two-body $\mathcal {B}$ (in $10^{-5}$)} &\multicolumn{1}{c}{} \\
 {Modes} & Scenario I~  &Scenario II~  &Exp~\cite{epjc78-1019} \\ \hline\hline
\hspace{1mm}$B^0 \to \eta_c K^+\pi^-(\rm NR)$           &$0.85^{+0.43+0.32+0.10}_{-0.41-0.26-0.15}$&$1.85^{+0.94+0.50+0.56}_{-0.59-0.24-0.31}$
                                                        &$5.90^{+1.23}_{-1.29}$\\
$B^0 \to \eta_c (K_0^*(1430)^0\to) K^+\pi^- $ &$1.69^{+0.71+0.22+0.42}_{-0.69-0.17-0.24}$&$4.75^{+2.10+1.17+1.05}_{-1.95-1.10-0.68}$
                                              &$14.50^{+3.36}_{-3.14}$\\
$ B_s^0    \to \eta_c K^-\pi^+(\rm NR)                  $&$0.03^{+0.01+0.01+0.00}_{-0.01-0.01-0.00}$&$0.09^{+0.04+0.03+0.02}_{-0.04-0.02-0.02}$&$\cdots$\\
$ B_s^0    \to \eta_c (\bar K_0^*(1430)^0\to) K^-\pi^+  $&$0.08^{+0.03+0.02+0.01}_{-0.02-0.01-0.01}$&$0.21^{+0.11+0.07+0.05}_{-0.08-0.05-0.03}$&$\cdots$\\
\hline
$ B^0    \to \eta_c(2S) K^+\pi^-(\rm NR)                   $&$0.17^{+0.07+0.03+0.04}_{-0.06-0.02-0.05}$&$0.41^{+0.22+0.12+0.11}_{-0.11-0.09-0.06}$&$\cdots$\\
$ B^0    \to \eta_c(2S) (K_0^*(1430)^0\to) K^+\pi^-    $&$0.56^{+0.28+0.17+0.12}_{-0.25-0.24-0.15}$&$0.79^{+0.41+0.24+0.17}_{-0.19-0.13-0.12}$&$\cdots$\\
$ B_s^0    \to \eta_c(2S) K^-\pi^+(\rm NR)                   $&$0.007^{+0.004+0.003+0.001}_{-0.002-0.001-0.002}$&$0.02^{+0.01+0.01+0.00}_{-0.01-0.01-0.00}$&$\cdots$\\
$ B_s^0    \to \eta_c(2S) (\bar K_0^*(1430)^0\to) K^-\pi^+    $&$0.02^{+0.01+0.01+0.00}_{-0.01-0.00-0.00}$&$0.04^{+0.02+0.01+0.01}_{-0.02-0.01-0.01}$&$\cdots$\\
\hline\hline
\end{tabular*}
\end{center}
\end{table}

\begin{table}  
\caption{ PQCD results for the branching ratios  of the $P$-wave resonances in the
$B^0_{(s)} \to \eta_c(1S,2S) K^{\pm} \pi^{\mp}$ decays together with experimental data~\cite{epjc78-1019,pdg2018}.
The theoretical errors are attributed to the variation of the Gegenbauer moments ($a_{1K^*}^{||}$, $a_{2K^*}^{||}$ and $a_{1s}^{\perp}$, $a_{1t}^{\perp}$),
the shape parameters $\omega_{B_{(s)}}$ in the wave function of the $B_{(s)}$ meson, and the hard scale $t$, respectively. }
\label{Tab:pwave}
\begin{tabular*}{15cm}{@{\extracolsep{\fill}}lll} \hline\hline
{\rm Modes}&{\rm Quasi-two-body ${\cal B}$ (in $10^{-5})$} &{\rm Exp  (in $10^{-5})$}\\ \hline
$ B^0    \to \eta_c(K^*(892)^0\to) K^+\pi^-        $ &$46.49^{+18.07+12.63+12.13}_{-14.80-9.67-8.63}$
                                                    &$\footnotemark[1]35\pm 5$~\cite{pdg2018}\\
$ B^0    \to \eta_c(K^*(1410)^0\to) K^+\pi^-       $ &$1.35^{+0.50+0.24+0.43}_{-0.40-0.23-0.26}$
                                                    &$1.20\pm0.90$~\cite{epjc78-1019}\\
$ B^0    \to \eta_c(K^*(1680)^0\to)   K^+\pi^-     $ &$1.44^{+0.63+0.34+0.55}_{-0.49-0.26-0.30}$
                                                    &$1.26^{+1.44}_{-1.51}$~\cite{epjc78-1019}\\
$ B_s^0    \to \eta_c(\bar K^*(892)^0\to) K^-\pi^+      $ &$2.13^{+0.89+0.66+0.54}_{-0.73-0.46-0.39}$&$\cdots$\\
$ B_s^0    \to \eta_c(\bar K^*(1410)^0\to) K^-\pi^+     $ &$0.07^{+0.02+0.02+0.02}_{-0.02-0.02-0.01}$&$\cdots$\\
$ B_s^0    \to \eta_c(\bar K^*(1680)^0\to)   K^-\pi^+   $ &$0.08^{+0.02+0.02+0.03}_{-0.02-0.02-0.02}$&$\cdots$\\
\hline
$ B^0    \to \eta_c(2S)(K^*(892)^0\to) K^+\pi^-        $ & $13.33^{+4.72+3.70+3.74}_{-3.51-2.96-2.08} $ & $\footnotemark[1]<26$~\cite{pdg2018} \\
$ B^0    \to \eta_c(2S)(K^*(1410)^0\to) K^+\pi^-       $&$0.24^{+0.11+0.06+0.10}_{-0.06-0.04-0.05}$&$\cdots$\\
$ B^0    \to \eta_c(2S)(K^*(1680)^0\to) K^+\pi^-       $&$0.11^{+0.04+0.02+0.04}_{-0.03-0.02-0.03}$&$\cdots$\\
$ B_s^0  \to \eta_c(2S)(\bar K^*(892)^0\to) K^-\pi^+        $&$0.60^{+0.20+0.21+0.16}_{-0.19-0.16-0.11}$&$\cdots$\\
$ B_s^0  \to \eta_c(2S)(\bar K^*(1410)^0\to) K^-\pi^+      $&$0.01^{+0.01+0.01+0.01}_{-0.00-0.00-0.00}$&$\cdots$\\
$ B_s^0  \to \eta_c(2S)(\bar K^*(1680)^0\to) K^-\pi^+      $&$0.008^{+0.002+0.002+0.002}_{-0.002-0.002-0.002}$&$\cdots$\\
\hline\hline
\end{tabular*}
\footnotetext[1]{ The experimental results are obtained by multiplying the relevant measured two-body branching ratios according to Eq.~(\ref{eq:def1}). }
\end{table}

\begin{table}  
\caption{ PQCD results  for the branching ratios  of the $D$-wave resonances in the
$B^0_{(s)} \to \eta_c(1S,2S) K^{\pm} \pi^{\mp}$ decays together with experimental data~\cite{epjc78-1019}.
The theoretical errors are attributed to the variation of the Gegenbauer moment $a_1^0$, the shape parameters $\omega_{B_{(s)}}$
in the wave function of the $B_{(s)}$ meson, and the hard scale $t$, respectively. }
\label{Tab:dwave}
\begin{tabular*}{15cm}{@{\extracolsep{\fill}}lll} \hline\hline
{\rm Modes}&{\rm Quasi-two-body ${\cal B}$ (in $10^{-5})$} &{\rm Exp  (in $10^{-5})$} \\ \hline
$ B^0    \to \eta_c (K_2^*(1430)^0\to) K^+\pi^-    $ &$3.98^{+1.24+0.59+0.11}_{-1.74-0.55-0.04}$
                                                    &$2.35^{+1.08}_{-1.29}$~\cite{epjc78-1019}\\
$ B_s^0    \to \eta_c (\bar K_2^*(1430)^0\to) K^-\pi^+  $ &$0.23^{+0.13+0.04+0.01}_{-0.10-0.05-0.01}$&$\cdots$\\
\hline
$ B^0    \to \eta_c(2S) (K_2^*(1430)^0\to) K^+\pi^-            $&$0.55^{+0.31+0.12+0.02}_{-0.24-0.09-0.01}$&$\cdots$\\
$ B_s^0  \to \eta_c(2S) (\bar K_2^*(1430)^0\to) K^-\pi^+       $&$0.04^{+0.02+0.01+0.01}_{-0.02-0.01-0.01}$&$\cdots$\\
\hline\hline
\end{tabular*}
\end{table}

By using Eqs.~(\ref{expr-br})--(\ref{br-momentum}), the decay amplitudes from the Appendix, and all of the input quantities,
the resultant branching ratios $\mathcal{B}$ and the available experimental results for the considered
$B_{(s)}^0 \to \eta_c K \pi$ decays involving the $S$-wave resonances are summarized in Table~\ref{Tab:swave},
while those for $P$- and $D$-wave resonances  are shown in Tables~\ref{Tab:pwave} and ~\ref{Tab:dwave}.
Since the charged $B$ meson decays differ from the neutral ones only in the lifetimes and the isospin factor in our
formalism, we can derive the branching ratios for the $B^+$ decay modes by multiplying those for the $B^0$ decay modes by the ratio $\tau_{B^+}/\tau_{B^0}$.

In our calculations for the various partial-wave resonances, the first uncertainty is induced by the Gegenbauer
moments in the $S$, $P$, and $D$-wave kaon-pion DAs, as aforementioned.
The second error comes from the variations of the shape parameter $\omega_{B_{(s)}}$ of the $B_{(s)}$ meson DA.
We adopt the value $\omega_B=0.40\pm0.04$ or $\omega_{B_s}=0.50\pm0.05$~GeV and vary its value within
a 10\% range, which is supported by intensive PQCD studies~\cite{prd63-074009,plb504-6}.
The last one is caused by the variation of the hard scale $t$ from $0.75t$ to $1.25t$ (without changing $1/b_i$), which
characterizes the  next-to-leading-order (NLO) effects in the PQCD approach.
In Tables~\ref{Tab:swave}, ~\ref{Tab:pwave}, and ~\ref{Tab:dwave}, it is shown that the main uncertainties
in our approach come from the Gegenbauer moments, which can reach  a total magnitude of about $60\%$ .
The scale-dependent uncertainty is less than $25\%$ due to the inclusion of the NLO vertex corrections.
The other possible errors from the uncertainties of $m_c$ and the Cabibbo-Kobayashi-Maskawa matrix elements are actually very small and can be safely neglected.

Combined with the Clebsch-Gordan Coefficients, we can write the relation
\begin{eqnarray}
|K\pi,I=\frac{1}{2} \rangle =\sqrt{\frac{1}{3}}|K^0\pi^0\rangle-{\sqrt{\frac{2}{3}}}|K^+\pi^-\rangle.
\end{eqnarray}
Isospin conservation is assumed for the strong decays of an $I=1/2$ resonance $K^{*0}$ to $K\pi$ when we compute the branching fraction of the quasi-two-body process $B\to \eta_c K^{*0} \to \eta_c K^+\pi^-$, namely,
\begin{eqnarray}
\frac{\Gamma(K^{*0} \to K^+\pi^-)}{\Gamma(K^{*0} \to K\pi)}=2/3, ~\quad
\frac{\Gamma(K^{*0} \to K^0\pi^0)}{\Gamma(K^{*0} \to K\pi)}=1/3. \label{eq:def2}
\end{eqnarray}
Therefore, the corresponding two-body branching fraction ${\cal B}(B \to \eta_c K^{*0})$ can be extracted directly from the quasi-two-body decay modes in Table~\ref{Tab:pwave} under the narrow-width approximation relation:
\begin{eqnarray}
\mathcal{B}(B \to \eta_c K^{*0}\to \eta_c K^+ \pi^- ) &=&
\mathcal{B}( B \to \eta_c K^{*0}) \cdot {\mathcal B}(K^{*0} \to K\pi)\cdot\frac{2}{3}.\label{eq:def1}
\end{eqnarray}
There already exist some results for $B_{(s)}^0 \to \eta_c K^*(892)^0$ in the two-body framework~\cite{prd71-114008,1909-10907,epjc58-245}.
One can see that the branching ratios of the quasi-two-body decay modes match well with the two-body analyses presented in Ref.~\cite{1909-10907} using the PQCD approach.
These results suggest that the PQCD factorization approach is suitable for describing
the quasi-two-body $B$ meson decays through analyzing various resonances by reconstructing $K\pi$ final states
and  reproducing the invariant mass spectra of Dalitz plots.

For the $S$-wave resonance $K_0^*(1430)^0$, it should be mentioned that two scenarios have been proposed to describe the scalar mesons above 1 GeV using the QCD sum rules method~\cite{prd73-014017,prd77-014034}.
In Scenario I, the $K^*_0(1430)^0$ is treated as the first excited state, while $a_0(980)$ and $f_0(980)$ are regarded as the lowest-lying states.
In Scenario II, we assume that $K_0^*(1430)^0$ is the lowest-lying resonance and the corresponding first excited states lie between (2.0--2.3) GeV.
Scenario II corresponds to the case that light scalar mesons are four-quark bound states.
The Gegenbauer moments $B_1=0.58\pm0.07$ and $B_3=-1.20\pm0.08$ are adopted in Scenario I, while $B_1=-0.57\pm0.13$ and $B_3=-0.42\pm0.22$ are adopted in Scenario II~\cite{prd77-014034}.
In this work, we consider two scenarios for the $S$-wave components and list the relevant results in Table~\ref{Tab:swave}.
One can see that the predicted $\cal B$ in Scenario I are always smaller than those in Scenario II.
This phenomenon is mainly caused by the different signs of the Gegenbauer moment $B_1$ in different scenarios, which indicates that there is a large cancellation in Scenario I.

From Table~\ref{Tab:swave}, one can see that our predictions for the branching ratios for the ${K}_0^*(1430)^0$ resonance and NR components in Scenario II are ${\cal B}_{K_0^*(1430)^0}=(4.75^{+2.62}_{-2.34})\times 10^{-5}$ and ${\cal B}_{\rm NR}=(1.85^{+1.20}_{-0.71})\times 10^{-5}$, respectively, which are a bit smaller than the experimental measurements within errors.
Anyway, as is well known, in contrast to the vector and tensor mesons the identification of scalar mesons is a long-standing puzzle.
It is difficult to deal with scalar resonances since some of them have wide decay widths, which cause a strong overlap between resonances and background.
Furthermore, the underlying structure of scalar mesons is not theoretically well established  (for a review, see Ref.~\cite{pdg2018}).
We hope that the situation can be improved using nonperturbative QCD tools including lattice QCD simulations.
Nonetheless, we define the PQCD prediction of the corresponding ratio for a more direct comparison with the available experimental data,
\beq
R_1^{\rm PQCD}&=&\frac{{\cal B}( B^0 \to \eta_c (K_0^*(1430)^0\to) K^+\pi^-)}
{ {\cal B}( B^0 \to \eta_c K^+\pi^-(\rm NR))}
= 2.56^{+1.98}_{-1.71},
\eeq
where the branching fraction of the $ B^0 \to \eta_c (K_0^*(1430)^0\to) K^+\pi^-$ decay is measured relative to that of the corresponding NR contributions by the LHCb Collaboration~\cite{epjc78-1019}:
\beq
R_1^{\rm LHCb}&=&\frac{{\cal B}( B^0 \to \eta_c (K_0^*(1430)^0\to) K^+\pi^-)}
{ {\cal B}( B^0 \to \eta_c K^+\pi^-(\rm NR))}
= 2.45^{+0.81}_{-0.68}.
\eeq
Our prediction is quite consistent with the LHCb measurement.
Combined analyses from the LHCb and Belle-II measurements for these decays in the near future could help us to further study the scalar resonances.

\begin{figure}[tbp]
\centerline{\epsfxsize=11cm \epsffile{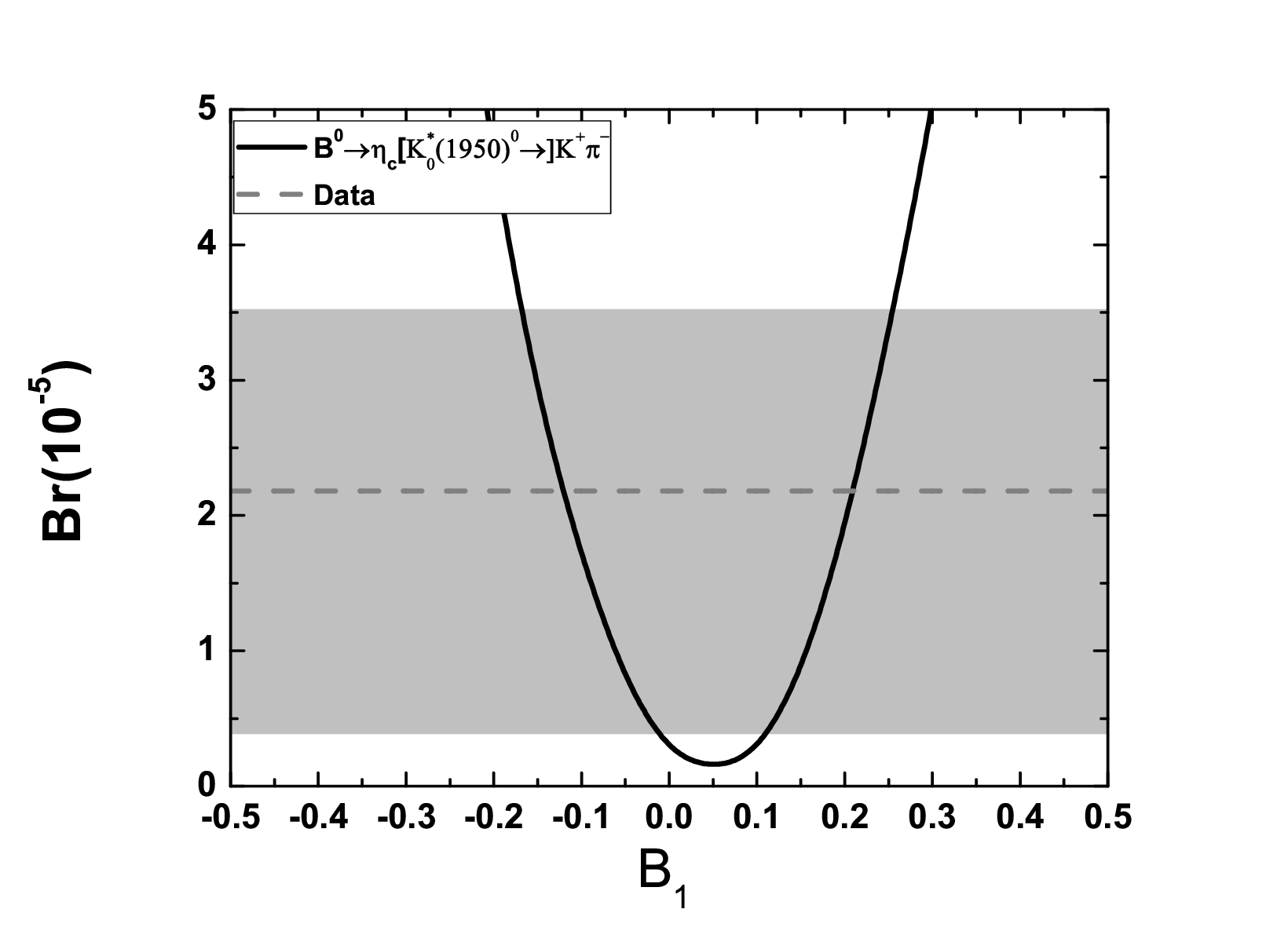}}
\caption{The branching fraction of the $B^0 \to \eta_c[K^*(1950)^0\to]K^+\pi^-$ decay as a function of the Gegenbauer moment $B_1$.
Shaded bands show the experimental uncertainties.}
\label{fig:fig2}
\end{figure}

For the phenomenological study of the scalar meson ${K}_0^*(1950)^0$, we still lack the distribution amplitudes of the ${K}_0^*(1950)^0$ state at present.
Fortunately, we are allowed to single out the ${K}_0^*(1950)^0$ component according to the kaon-pion DAs.
In Fig.~\ref{fig:fig2}, we plot the variation of the branching fraction with the first odd Gegenbauer moment $B_1$ for the $B^0 \to \eta_c (K_0^*(1950)^0\to) K^+\pi^-$ decay mode, as well as the experimental data ${\cal B}_{\rm exp}=(2.18^{+1.32}_{-1.79})\times 10^{-5}$~\cite{epjc78-1019}.
One can see that the theoretical prediction of the branching ratio shows a strong dependence on the variation of  $B_1$.
Combined with the available data, we can roughly determine that the possible range of first odd Gegenbauer moment is from $-0.15$ to $0.05$ or $0.10$ to $0.25$, which should be examined both theoretically and experimentally in the future.

\begin{figure}[tbp]
\centerline{\epsfxsize=11cm \epsffile{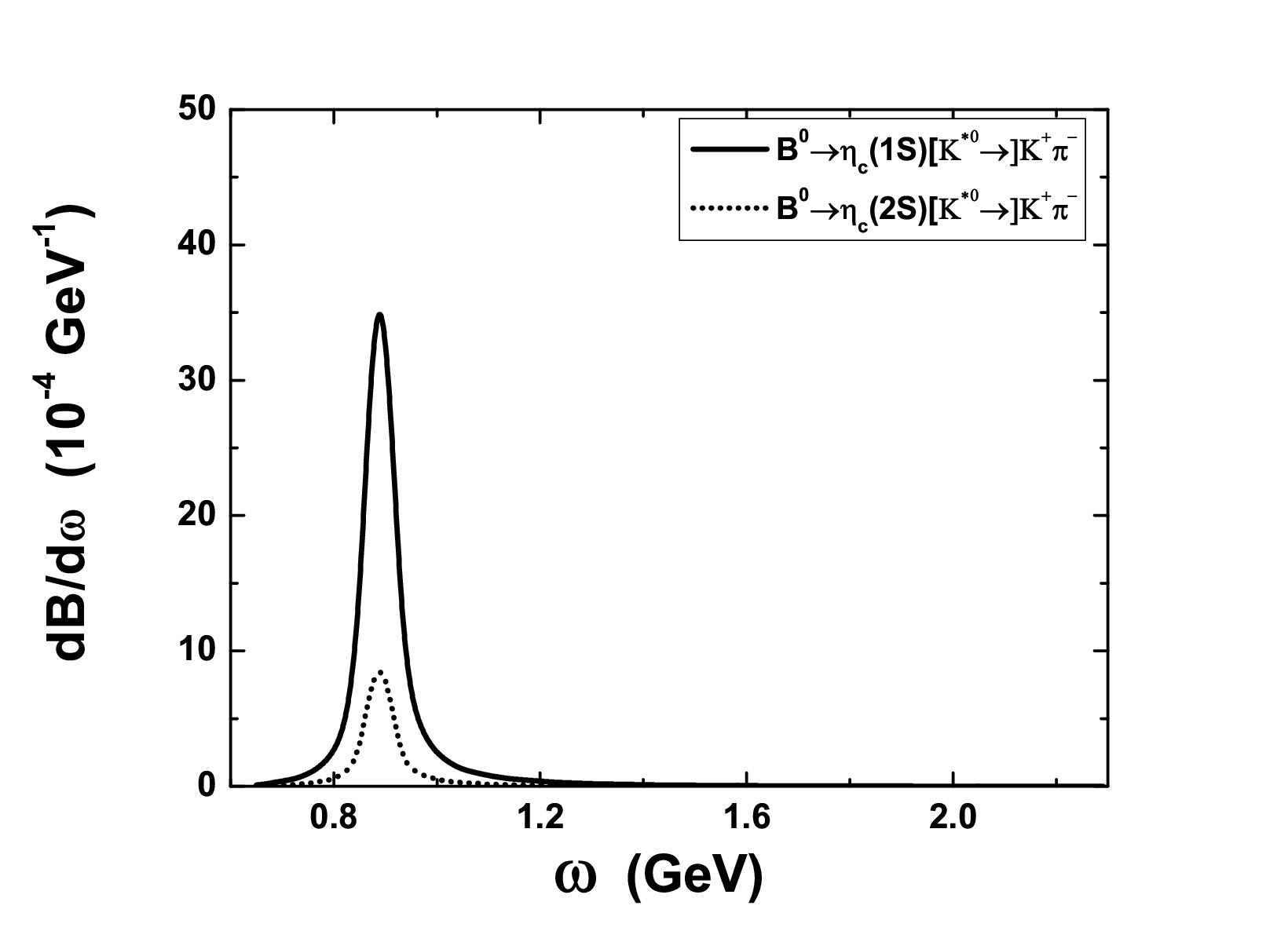}}
\caption{Differential branching ratio for the $B^0 \to \eta_c(1S,2S)[K^{*}(892)^0\to]K^+\pi^-$ decays.}
\label{fig:fig3}
\end{figure}

In Fig.~\ref{fig:fig3}, we show the $\omega$ dependence of the differential decay ratas
$d{\cal B}(B^0 \to \eta_c K^+\pi^-)/d\omega$ (the solid curve) and $d{\cal B}(B^0 \to \eta_c(2S)K^+\pi^-)/d\omega$ ( short-dotted curve) after the inclusion of the $P$-wave resonance $K^*(892)^0$, which exhibit a peak at the $K^*(892)^0$ meson mass.
For the considered decay modes $B_{(s)}^0 \to \eta_c(1S,2S)K\pi$, the dynamical limit on the value of the invariant  mass $\omega$ is $(m_K+m_{\pi})\leq \omega \leq (m_{B_{(s)}}-m_{\eta_c(1S,2S)})$.
Although $m_{K^*(1680)^0}>(m_{B}-m_{\eta_c(2S)})$, the resonance $K^*(1680)^0$ can still contribute to the $B_{(s)}^0\to\eta_c(2S)K \pi$ decay
due to its wide width ($\Gamma_{K^*(1680)^0}=322$ MeV).
It is shown that the main portion of the differential branching ratio
 lies in the region around the resonance in Fig.~\ref{fig:fig3}, as expected.
For $B^0 \to \eta_c(1S) (K^*(892)^0 \to)K^+\pi^-$ decay, the central values of the branching ratio ${\cal B}$ are $23.36\times10^{-5} $ and $34.46\times 10^{-5}$ when the integration over $\omega$ is limited  to the range
$\omega=[m_{K^*}-0.5\Gamma_{K^*}, m_{K^*}+0.5\Gamma_{K^*}]$ or
$\omega=[m_{K^*}-\Gamma_{K^*}, m_{K^*}+\Gamma_{K^*}]$, respectively, which amounts to
$50\%$ and $74\%$ of the total branching ratio ${\cal B}=46.49\times10^{-5}$ as listed in Table~\ref{Tab:pwave}.

From Table~\ref{Tab:pwave}, one can see that our PQCD prediction for the branching ratio of the
$B^0 \to \eta_c(1S)K^*(892)^0 \to \eta_c K^+\pi^- $ decay is ${\cal B}=(46.49^{+25.16}_{-19.67})\times 10^{-5}$, the central value of which is a little larger than that in PDG2018: $(35\pm 5)\times 10^{-5}$~\cite{pdg2018}.
The measurements from the Belle~\cite{Prl90-071801}, {\it BABAR}~\cite{PRD76-092004,PRD78-012006},
and LHCb~\cite{epjc78-1019} collaborations, as well as the average value from HFLAV~\cite{hf2019}, are the following:
\beq
{\calb }(B^0 \to \eta_c (K^*(892)^0\to)K^+\pi^-)=
\left\{\begin{array}{ll}
(108^{+42}_{-46})\times 10^{-5}         & {\rm Belle}~[2], \\
(53^{+28}_{-19})\times 10^{-5}         & {\rm BABAR}~[3],\\
(38\pm 7)\times 10^{-5}                & {\rm BABAR}~[4],\\
(29.5^{+3.8}_{-4.7})\times 10^{-5}      & {\rm LHCb} ~[5],\\
(41.3\pm 6.6)\times 10^{-5}             & {\rm HFLAV} ~[80].\\
\end{array} \right. \label{eq:ratio3}
\eeq
The data $(41.3\pm 6.6)\times 10^{-5}$ from $\rm HFLAV$~\cite{hf2019} is obtained by multiplying the relevant measured two-body branching ratio according to the Eq.~(\ref{eq:def1}).
One can see that the central values of the measured branching ratio for the $K^*(892)^0$ resonance from different experiments vary in the wide range $(29-108)\times 10^{-5}$, while the HFLAV world average of the measured values from the Belle and {\it BABAR} collaborations~\cite{Prl90-071801,PRD76-092004,PRD78-012006} leads to
$(41.3\pm 6.6)\times 10^{-5}$, which is in good agreement with our prediction.

For the $B^0 \to \eta_c(2S) (K^*(892)^0 \to) K^+\pi^-$ decay, the {\it BABAR} Collaboration has measured an upper limit on the branching ratio ${\cal B}( B^0 \to \eta_c(2S) (K^*(892)^0 \to) K^+\pi^-) <26 \times 10^{-5}$ at the $90\%$ confidence level~\cite{PRD78-012006}.
The PQCD prediction of ${\cal B}( B^0 \to \eta_c(2S) (K^*(892)^0\to) K^+\pi^-)=(13.33^{+7.07}_{-5.04}) \times 10^{-5}$ agrees with the limit.
Meanwhile, one can see in Fig.~\ref{fig:fig3} that the branching fractions of $B^0 \to \eta_c(2S)(K^{*0}\to)K^+\pi^-$ decays are always smaller than those for $B^0 \to \eta_c(1S)(K^{*0}\to)K^+\pi^-$ decays, which is mainly induced by the difference between the DAs of the $\eta_c(2S)$ and $\eta_c(1S)$ mesons: the tighter phase space and the smaller decay constant of $\eta_c(2S)$ meson result in the suppression.

From the numerical results given in Table~\ref{Tab:pwave}, we obtain the relative ratio $R_2$ between the branching ratio of $B$ meson decays involving $\eta_c(2S)$ and $\eta_c(1S)$ and the resonance $K^*(892)^0$,
\beq
R_2(K^*(892)^0)&=&\frac{{\cal B}( B^0 \to \eta_c(2S)(K^*(892)^0\to) K^+\pi^-)}
{ {\cal B}(B^0\to \eta_c(1S)(K^*(892)^0\to) K^+\pi^-)}
= 0.29^{+0.21}_{-0.15}
\eeq
which can be tested by the forthcoming LHCb and Belle-II experiments.

The branching ratios of the considered $D$-wave resonance are presented in Table~\ref{Tab:dwave}.
We emphasized that our predictions of these decay channels are only rough estimates.
Although there is not enough data at present, the calculated value ${\cal B}(B^0 \to \eta_c(1S)(K_2^*(1430)^0\to) K^+\pi^-)=(3.98^{+1.24}_{-1.74})\times 10^{-5}$ is compatible  with the measurement $(2.35^{+1.08}_{-1.29})\times 10^{-5}$~\cite{epjc78-1019} within the large errors.
Future experimental measurements with high precision can provide us with a better understanding of the properties of the tensor resonances.

For all of the $B^0_s \to \eta_c(1S,2S)K\pi$ decays, such decay modes can be theoretically related to the corresponding $B^0$ decays since they have identical topologies and similar kinematic properties in the limit of $SU(3)$ flavor symmetry.
At the quark level, the $B^0$ and $B^0_s$ decays correspond to the $b\to c{\bar c}s$ and $b\to c{\bar c}d$ transitions, respectively.
The relative ratios of the branching fractions for $B^0_s$ and $B^0$ decay modes are dominated by the Cabibbo suppression factor of $|V_{cd}|^2/|V_{cs}|^2\sim \lambda^2$ under the naive factorization approximation.
It is reasonable to see that the branching fractions of the $B^0_s$ decays are smaller than those of the corresponding $B^0$ decays.
Though the $B^0_s$ channels have relatively small branching ratios, some of them can be potentially measurable at future experiments.

\section{CONCLUSION}

In this work, by introducing the kaon-pion DAs, we studied the quasi-two-body decays $B^0_{(s)} \to \eta_c(1S,2S) (K^{*0}\to) K\pi$
in the PQCD approach, in which the kaon-pion invariant mass spectra are dominated by the $K_0^*(1430)^0,K_0^*(1950)^0,K^*(892)^0,
K^*(1410)^0,K^*(1680)^0$, and $K_2^*(1430)^0$ resonances.
These six resonances fall into three partial waves according to their spins, namely, $S,P$, and $D$-wave states.
The contributions from each partial wave can be parametrized into the corresponding time-like form factors involved in the kaon-pion DAs.
The $K\pi$ $S$-wave component at low mass is described by the LASS line shape, while the time-like form factors of other resonances are
modeled by the relativistic BW function.

It has been shown that our predictions of the branching ratios for most of the considered $B^0 \to \eta_c(1S) (K^{*0}\to) K^+\pi^-$ decays
are in good agreement with the existing data within the errors.
For the $B^0 \to \eta_c(1S) (K_0^*(1430)^0\to) K^+\pi^-$ decay, although there exists a clear difference between the central value of the
PQCD calculation for ${\cal B}_{K_0^*(1430)^0}$, ${\cal B}_{\rm NR}$ and  the measured ones, they are still consistent within three standard
deviations due to the large experimental errors, which should be examined by forthcoming experiments.
The new ratio $R_2(K^*(892)^0)$ among the branching ratios of the considered decay modes has been defined and will be confronted with  future measurements.

\begin{acknowledgments}

Many thanks to Hsiang-nan Li and Wen-Fei Wang for valuable discussions.
This work was supported by the National Natural Science Foundation of China under the No.~11947013, No.~11605060, No.~11775117, and No.~11547020.
Y.~L. is also supported by the Natural Science Foundation of Jiangsu Province under Grant No.~BK20190508 and the Research Start-up Funding of Nanjing
Agricultural University.
Z.~R. is supported in part by the Natural Science Foundation of Hebei Province under Grant No.~A2019209449.

\end{acknowledgments}
\appendix

\section{Decay amplitudes}
The total decay amplitudes for the considered decay modes $B^0_{(s)} \to \eta_c K\pi$ in this work are given as follows:
\begin{eqnarray}
\mathcal{A}(B^0_{(s)}\to \eta_c K\pi)&=&\frac{G_F}{\sqrt{2}}\Big\{V^*_{cb}V_{cs(cd)}
\Big [(C_1+\frac{1}{3}C_2)\mathcal{F}^{LL}+C_2\mathcal{M}^{LL} \Big]\nonumber\\
&-&V^*_{tb}V_{ts(td)}\Big [(C_3+\frac{1}{3}C_4+C_9+\frac{1}{3}C_{10})\mathcal{F}^{LL}
+(C_5+\frac{1}{3}C_6+C_7+\frac{1}{3}C_{8})\mathcal{F}^{LR}\nonumber\\
&+&(C_4+C_{10})\mathcal{M}^{LL}+(C_6+C_8)\mathcal{M}^{SP}\Big ]\Big\},
\end{eqnarray}
where $G_F=1.16639\times 10^{-5}$ GeV$^{-2}$ is the Fermi coupling constant and the $V_{ij}$'s are the Cabibbo-Kobayashi-Maskawa matrix elements.
The superscripts $LL$, $LR$, and $SP$ refer to the contributions from $(V-A)\otimes(V-A)$, $(V-A)\otimes(V+A)$, and $(S-P)\otimes(S+P)$ operators, respectively.
The explicit amplitudes $\mathcal{F(M)}$ from the factorizable (nonfactorizable) diagrams in Fig.~\ref{fig:fig1} can be obtained straightforwardly by replacing the twist-2 or twist-3 DAs of the $\pi\pi$ and $KK$ systems with the corresponding twists of the $K\pi$ ones in Eqs.~(\ref{eq:phi1s})--(\ref{eq:phi3s}), (\ref{eq:phi1p})--(\ref{eq:phi3p}), and (\ref{eq:phi1d})--(\ref{eq:phi3d}), since the kaon-pion distribution amplitudes considered in this work [Eqs.~(\ref{eq:fuliye2}), (\ref{eq:pwavekpi}), and (\ref{eq:dwavekpi})] have the same Lorentz structure as the two-pion (kaon) ones in Refs.~\cite{epjc76-675,epjc79-792}.

The Wilson coefficients $C_i$ are evaluated at the corresponding scale $t$. At $m_W$ scale, the Wilson coefficients at the NLO level can be written in the following form (as in Ref.~\cite{rmp681125}):
\begin{eqnarray}
C_{1}(m_W)&=&\frac{11}{2}\frac{\alpha_s(m_W)}{4\pi},\nonumber\\
C_{2}(m_W)&=&1-\frac{11}{6}\frac{\alpha_s(m_W)}{4\pi}-\frac{35}{8}\frac{\alpha_{em}}{4\pi},\nonumber\\
C_{3}(m_W)&=&-\frac{\alpha_s(m_W)}{24\pi}\left[ E_0(x)-\frac{2}{3}\right]+\frac{\alpha_{em}}{6\pi}\frac{1}{{\rm sin}^2\theta_W}\left[ 2B_0(x)+C_0(x)\right],\nonumber\\
C_{4}(m_W)&=&\frac{\alpha_s(m_W)}{8\pi}\left[ E_0(x)-\frac{2}{3}\right],\nonumber\\
C_{5}(m_W)&=&-\frac{\alpha_s(m_W)}{24\pi}\left[ E_0(x)-\frac{2}{3}\right],\nonumber\\
C_{6}(m_W)&=&\frac{\alpha_s(m_W)}{8\pi}\left[ E_0(x)-\frac{2}{3}\right],\nonumber\\
C_{7}(m_W)&=&\frac{\alpha_{em}}{6\pi}\left[ 4C_0(x)+D_0(x)-\frac{4}{9}\right],\nonumber\\
C_{9}(m_W)&=&\frac{\alpha_{em}}{6\pi}\left[ 4C_0(x)+D_0(x)-\frac{4}{9}+\frac{1}{{\rm sin}^2\theta_W}(10B_0(x)-4C_0(x))\right],\nonumber\\
C_{i}(m_W)&=&0,i=8,10,
\end{eqnarray}
where the relevant Inami-Lim functions $B_0(x), C_0(x), D_0(x)$, and $E_0(x)$ \cite{Inami} are
\begin{eqnarray}
B_0(x)&=&\frac{1}{4}\left( \frac{x}{1-x}-\frac{x{\rm ln}x}{(x-1)^2}\right),\nonumber\\
C_0(x)&=&\frac{6x-x^2}{8(1-x)}-\frac{(3x^2+2x){\rm ln}x}{8(x-1)^2},\nonumber\\
D_0(x)&=&\frac{-25x^2+19x^3}{36(1-x)^3}-\frac{(8-32x+54x^2-30x^3+3x^4){\rm ln}x}{18(x-1)^4},\nonumber\\
E_0(x)&=&\frac{18x-11x^2-x^3}{12(1-x)^3}-\frac{(4-16x+9x^2){\rm ln}x}{6(x-1)^4},
\end{eqnarray}
with $x=m_t^2/m_W^2$.
In the region $m_b<t<m_W$,  we evaluate the Wilson coefficients at $t$ scale using the following renormalization group equation
\begin{equation}
\textbf{C}(t)=U(t,m_W)\textbf{C}(m_W),
\end{equation}
where $\textbf{C}(m_W)=(C_1(m_W),...,C_{10}(m_W))^T$ and $U(t,m_W)$ is  the renormalization group running
matrix at NLO level  ( for details, see Ref.~\cite{rmp681125}).
The Wilson coefficients $C_i(t)$ evaluated at the scale $t=m_b=4.8$ GeV (as given, for example, in Ref.~\cite{epjc59-49}) are as follows:
\begin{eqnarray}\label{eq:cib}
C_1&=&-0.17474,\quad C_2=1.07737, \nonumber\\
C_3&=&0.01249,\quad C_4=-0.03304, \nonumber\\
C_5&=&0.00942,\quad C_6=-0.03929, \nonumber\\
C_7&=&-0.00003,\quad C_8=0.00023, \nonumber\\
C_9&=&-0.00999,\quad C_{10}=0.00201.
\end{eqnarray}
If the scale $ t < m_b$, we can evaluate the Wilson coefficients at $t$ scale using the evolution equation $\textbf{C}(t)=U(t,m_b)\textbf{C}(m_b)$,
where $\textbf{C}(m_b)=(C_1(m_b),...,C_{10}(m_b))^T$ are given in Eq.~(\ref{eq:cib}).


\end{document}